\documentclass[10pt,preprint]{aastex}

\usepackage{booktabs}
\usepackage{natbib}

\shorttitle{SN 2010jl}
\shortauthors{Andrews et al.}

\begin{document}

\title{Evidence for Pre-Existing Dust in the Bright Type IIn SN 2010jl}

\author{J.E. Andrews\altaffilmark{1}, Geoffrey C. Clayton\altaffilmark{1}, R. Wesson\altaffilmark{2}, B.E.K. Sugerman\altaffilmark{3}, M.J. Barlow\altaffilmark{2}, J. Clem\altaffilmark{1}, B. Ercolano\altaffilmark{4},  J. Fabbri\altaffilmark{2}, J.S. Gallagher\altaffilmark{5}, A. Landolt\altaffilmark{1}, M. Meixner\altaffilmark{6},  M. Otsuka\altaffilmark{6}, D. Riebel\altaffilmark{7}, and D.L. Welch\altaffilmark{8}}
\altaffiltext{1}{Department of Physics and Astronomy, Louisiana State University, 202 Nicholson Hall, Baton Rouge, LA 70803; jandrews@phys.lsu.edu, landolt@phys.lsu.edu, gclayton@fenway.phys.lsu.edu, jclem@phys.lsu.edu}
\altaffiltext{2}{Department of Physics and Astronomy, University College London, Gower Street, London WC1E 6BT, UK; mjb@star.ucl.ac.uk, jfabbri@star.ucl.ac.uk, rwesson@star.ucl.ac.uk}
\altaffiltext{3}{Department of Physics and Astronomy, Goucher College, 1021 Dulaney Valley Rd., Baltimore, MD 21204; ben.sugerman@goucher.edu}
\altaffiltext{4}{Ludwig-Maximilians-Universitaet, University Observatory Munich (USM) ,Scheinerstr 1, D-81679, Muenchen Germany ; ercolano@usm.uni-muenchen.de}
\altaffiltext{5}{Department of Mathematics, Physics, and Computer Science, Raymond Walters Collge, 9555 Plainfield Rd., Blue Ash, OH 45236; gallagjl@ucmail.uc.edu}
\altaffiltext{6}{Space Telescope Science Institute, 3700 San Martin Drive, Baltimore, MD 21218; meixner@stsci.edu, otsuka@stsci.edu}
\altaffiltext{7}{Department of Physics and Astronomy, Johns Hopkins University, 3400 N. Charles Street, Baltimore, MD 21218; driebel@pha.jhu.edu}
\altaffiltext{8}{Department of Physics and Astronomy, McMaster University, Hamilton, Ontario, L8S 4M1 Canada; welch@physics.mcmaster.ca}

\begin{abstract}
SN 2010jl was an extremely bright, Type IIn SNe which showed a significant IR excess no later than 90 days after explosion. We have obtained Spitzer 3.6 and 4.5 $\mu$m and JHK observations of SN 2010jl $\sim$90 days post explosion.  Little to no reddening in the host galaxy indicated that the circumstellar material lost from the progenitor must lie in a torus inclined out of the plane of the sky. The likely cause of the high mid-IR flux is the reprocessing of the initial flash of the SN by pre-existing circumstellar dust. Using a 3D Monte Carlo Radiative Transfer code, we have estimated that between 0.03-0.35 M$_{\sun}$ of dust exists in a circumstellar torus around the SN located 6 $\times$ 10$^{17}$ cm away from the SN and inclined between 60-80$^{\circ}$ to the plane of the sky. On day 90, we are only seeing the illumination of approximately 5$\%$ of this torus, and expect to see an elevated IR flux from this material up until day $\sim$ 450. It is likely this dust was created in an LBV-like mass loss event of more than 3 M$_{\sun}$, which is large but consistent with other LBV progenitors such as $\eta$ Carinae. 

\end{abstract}


\section{Introduction}
SN 2010jl was the brightest core collapse supernova (CCSN) of 2010, with a peak unfiltered brightness
of 12.9 mag, and a corresponding peak absolute magnitude of roughly -20 \citep{2010CBET.2532....1N}. A spectrum taken on 2010 November 5 was classified as a Type IIn \citep{2010CBET.2536....1B}, a particular class of Type II SNe which show narrow ($\sim$100 km s$^{-1}$) and often intermediate ($\sim$1000 km s$^{-1}$) lines in H and He along with normal broad (10000 km s$^{-1}$) lines due to the expansion of the ejecta \citep{1990MNRAS.244..269S}.  The narrow emission lines in their spectra are attributed to the ionization of the pre-existing circumstellar material (CSM) which has been excited by the initial flash of the supernova. Intermediate lines can arise at early times due to the high optical depth in the CSM, which can cause multiple scatterings from thermal electrons which results in the broadening of the narrow lines  \citep{2010ApJ...709..856S}. It is believed the progenitors of Type IIn SNe,  which constitute 9-10$\%$ of the total population of CCSNe \citep{2011MNRAS.412.1522S,2011MNRAS.412.1441L,2009MNRAS.395.1409S},  lose 10$^{-4}$ to 10$^{-2}$ M$_{\sun}$ yr$^{-1}$ \citep{2011arXiv1104.5012F,2010arXiv1010.2689K} orders of magnitude more material than normal Type II progenitors. Pre-discovery Hubble Space Telescope (HST) images indicate that the progenitor of SN 2010jl was likely a massive star, $>$30 M$_{\sun}$, which could have been a luminous blue variable (LBV) in outburst phase \citep{2011ApJ...732...63S}.

SN 2010jl is only the second CCSN progenitor with a mass greater than 17 M$_{\sun}$ detected, the other being the LBV progenitor of SN 2005gl \citep{2009Natur.458..865G,2007ApJ...656..372G}. This is important for the study of dust production around CCSNe since the SNe from less massive progenitors (8-17 M$_{\sun}$) only seem to be producing 10$^{-2}$ - 10$^{-4}$ M$_{\sun}$ of dust \cite[for example]{2003MNRAS.338..939E,2006Sci...313..196S,2007ApJ...665..608M,2009ApJ...704..306K,2010ApJ...715..541A}. This is far less than the $\sim$ 0.1-1 M$_{\sun}$ of dust estimated per CCSNe to explain the large amounts of dust seen in galaxies of z $>$ 6.  This has led to speculation that more massive progenitors existed among the Population III stars of high-z galaxies, which may produce more dust \citep[for example]{2010ApJ...713....1C}.  To date though, there is indirect evidence that some nearby dust producing CCSNe have progenitors massive than 17 M$_{\sun}$, for example SN 2007it \citep{2011ApJ...731...47A}, and are also only producing dust masses of $\sim$ 10$^{-3}$ M$_{\sun}$.  This may indicate that the amount of dust produced is independent of progenitor mass. The study of dust formation in SN 2010jl, which likely has a massive progenitor, will provide an opportunity to explore this relationship more fully.

Due to the nature of Type IIn SNe, we expect pre-existing CSM gas and dust to exist around the SN. Therefore to distinguish newly formed dust from dust that was already present, it is imperative to take early-time observations to measure the contribution from pre-existing dust to serve as a baseline for later times when conditions may be favorable for new dust condensation.  Studying the pre-existing dust around CCSNe can also be essential to understanding the progenitor properties. Continued observations of the interaction between the SN and the surrounding CSM material over the first few years can reveal valuable information about densities, velocities, and over all mass-loss history of the progenitor as well as progenitor mass itself \citep[and references therein]{2011arXiv1104.5012F,2009ApJ...695.1334S}.   This paper on SN 2010jl presents a spectral energy distribution (SED) using optical, near- and mid- infrared photometry of SN 2010jl obtained $\sim$ 90 days after explosion and estimates of the mass, size, and age of the circumstellar dust.

\section{Observations}

Although the discovery date of SN 2010jl was 2010 November 3.5 \citep{2010CBET.2532....1N}, comparison of the optical spectra with other similar SNe \citep{2010CBET.2539....1Y} as well as pre-discovery images \citep{2010arXiv1012.3461S} indicate that the explosion date was likely early October. An ASAS image taken on 2010 October 9.6 shows the SN with a V magnitude of 13.79 \citep{2010arXiv1012.3461S}.  This is consistent with the GELATO modeling of SN 2010jl to SN 2006tf, which indicated that it was about a month past explosion \citep{2010CBET.2539....1Y}, and has led us to adopt 2010 October 10 (JD 2455480) as the date of explosion throughout this paper.  It is located in the interacting galaxy UGC 5189A, for which throughout this paper we adopt the distance of 48.9 Mpc \citep{2011ApJ...732...63S}. 
We have also assumed a Galactic foreground reddening of E(B-V) = 0.027 from \citet{1998ApJ...500..525S}, and are assuming zero reddening in the host galaxy based on the previous estimates for SN 2010jl by  \citet{2011A&A...527L...6P} and \citet{2011ApJ...732...63S}.

\begin{figure}[h]
   \centering
   \includegraphics[width=5.5in]{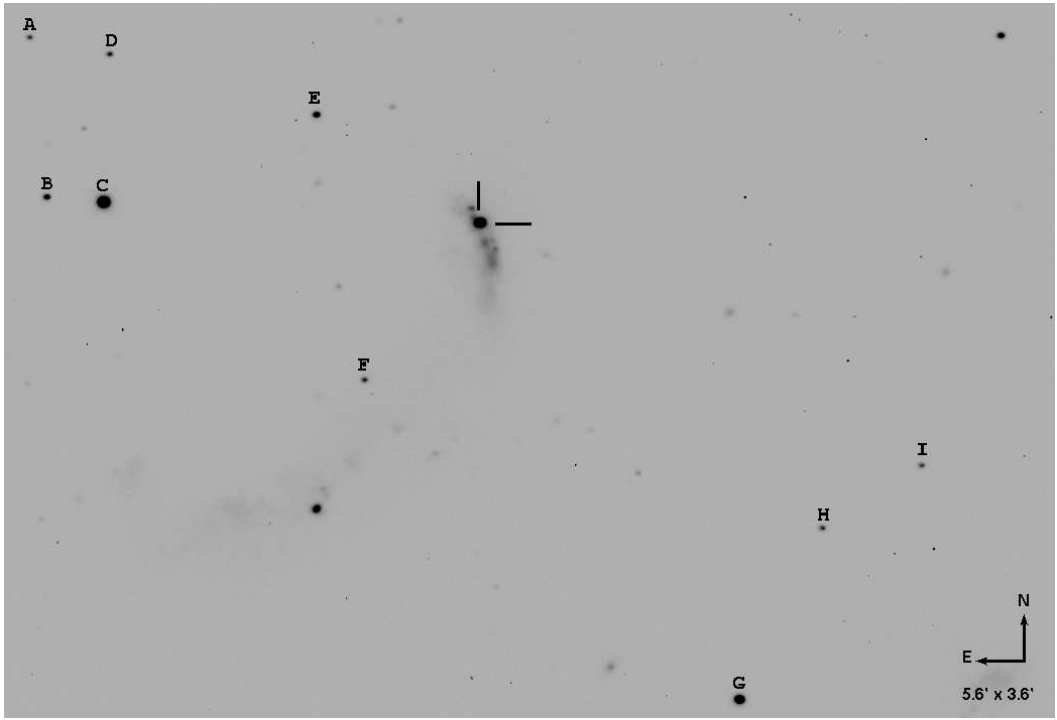} 
   \caption{V-band finder chart for tertiary standards for the field around UGC 5189 on 5 April 2011. Magnitudes are listed in Table 1.  }
   \label{fig:standards}
\end{figure}

A UVBRI photometric sequence of the UGC 5189A field (listed in Table 1 and shown in Figure \ref{fig:standards}) was taken on the night of 2010 December 5 (JD 2455504) at the KPNO 2.1m telescope.  These tertiary standards were calculated using the same methods as \citet{2010ApJ...715..541A,2011ApJ...731...47A}. The processing of all images to remove instrumental signatures was accomplished using the standard techniques of subtracting a median filtered bias frame and dividing by master twilight flat field images for each filter.  Also, the fringing effects in the I-band exposures were removed by scaling and subtracting a master fringe frame from the program images.  In order to transform the instrumental magnitudes to the standard system, between 30 and 35 different stars (depending on filter) were observed from the lists of  \citet{2009AJ....137.4186L} in addition to the SN 2010jl field. These stars were observed both near zenith and at a high airmass ($\sim$1.75) and were selected due to their color ranges.  Thus, they provided a viable sample of stars from which to derive accurate color and extinction coeffecients as well as magnitude zero-points for each filter.  The resulting RMS scatter of the residuals from the transformations revealed that the photometry for selected stars in the SN 20010jl field derived from this night's observations is accurate at the 1-2$\%$ level depending on filter.
BVRI photometry was obtained either from observations from the KPNO 1.2m (Table 2), the literature \citep{2010arXiv1012.3461S} or via the AAVSO data server and is plotted in Figure 2. The data shown in the plot have not been corrected for foreground extinction.  For the SED we used an observation from 2011 January 3 (JD 2455563), day 88, submitted to the AAVSO by contributer Etienne Morelle (Observations from the AAVSO International Database, 2010, private communication).

JHK$_{s}$ observations of SN 2010jl were taken on 2011 January 21 (JD 245583), or day 108, with the WHIRC camera on the WIYN telescope  \citep{2008SPIE.7014E..96M}. The detector offers a pixel scale of 0.098\arcsec pix$^{-1}$, and a FOV of  3\arcmin x 3\arcmin.  All images were taken in dither mode and aligned and stacked prior to flat fielding and bias correction. Eleven frames of 120s were taken in the J band, 12 frames of 120s in the H band, and 13 frames of 90s in the K band were taken of the SN 2010jl field. The transformation from instrumental to JHK$_{s}$ magnitudes was accomplished using 2MASS stars located in the frame for comparison photometry. Uncertainties were calculated by adding in quadrature the uncertainties associated with the 2MASS comparison stars and the uncertainties in the PSF photometry. The photometry is listed in Table 3 and shown in Figure 3.

The Spitzer IRAC (3.6 and 4.5 $\mu$m) images taken on 2011 January 5 (JD 2455565), corresponding to day 90, were mosaicked and resampled using standard MOPEX procedures to improve photometric quality. Pre-explosion IRAC images of UGC 5189A were also available in the Spitzer archive from 27 December 2007 (Program 40301, PI Fazio) which were used to subtract from the SN 2010jl images to get accurate photometry. PSF photometry was performed using the position specific PRF images.  Table 3 contains the measured mid-IR fluxes obtained from Spitzer. Figure 3 presents the SED of SN 2010jl on $\sim$ day 90 (88 for BVRI and 108 for JHK). Statistical uncertainties presented in the plot represent 1$\sigma$ errors.   The data shown in Figure 3 have been corrected for foreground extinction of E(B-V) = 0.027  \citep{1998ApJ...500..525S}.

\section{Discussion}

The SED of SN 2010jl, shown in Figure 3, shows an IR excess on day 90 which indicates the presence of warm dust around the SN. The IR excess  could be explained by two possible scenarios, one being that pre-existing CSM dust existing beyond the evaporation radius is reprocessing the initial flash of the SN into the infrared. Another possibility, although less likely, is that there is already dust forming in a cool dense shell (CDS) that has formed between the forward and reverse shocks created between the ejecta and the pre-existing CSM gas interaction \citep{1994ApJ...420..268C,2004MNRAS.352..457P}.  SN 2006jc \citep{2008ApJ...680..568S} and SN 2005ip \citep{2009ApJ...695.1334S, 2009ApJ...691..650F} formed dust at day 50 and 75 respectively, due to interaction with the CSM.  This newly formed dust would also create an IR excess.  Without optical spectra taken at the same epoch to look for the emergence of asymmetries, we cannot  clearly distinguish between pre-existing and newly formed dust.

The progenitor star of a CCSN is expected to have lost massive amounts of gas and dust prior to explosion, which will exist in the CSM.  When the star becomes a supernova, the initial UV flash from the explosion will evaporate the dust grains a certain distance from the SN, dependent on the initial luminosity. According to \citet{1983ApJ...274..175D,1985ApJ...297..719D}, and \citet{2009ApJ...691..650F}, a SN with an initial luminosity of 1 $\times$ 10$^{10}$ L$_{\sun}$ will clear a cavity with a radius, r$_{evap}$ = 6  $\times$ 10$^{16}$ cm if the CSM is carbon rich, or r$_{evap}$ = 3 $\times$ 10$^{17}$ cm if the CSM is oxygen rich.  Assuming the maximum luminosity of SN 2010jl was 3  $\times$ 10$^{43}$ erg s$^{-1}$ (M$_{v}$=-20), this is roughly 10$^{10}$  L$_{\sun}$, we can expect a cavity of cleared material around the SN to have a radius of approximately the same size.

Using black body fits (shown in Figure 3), we made initial estimates of R$_{in}$ (the inner radius of the shell), luminosity and temperature to constrain the model. These yielded a dust temperature (T$_{d}$) of roughly 750 K, ejecta temperatures (T$_{ej}$) of 7500 K , and an R$_{in}$ = 1.3  $\times$ 10$^{17}$ cm for this epoch.  The blackbody temperature and radius of the IR component of SN 2010jl suggests pre-existing dust, since new dust formed in the CDS prior to day 90 would be expected to be much hotter, $\sim$1600 K, as was the case in SN 2006jc and SN 2005ip and much closer to the expansion center, since the ejecta on day 90 would have traveled a maximum of 1.1 $\times$ 10$^{16}$ cm assuming a constant expansion of 14 000 km s$^{-1}$ \citep{2010CBET.2536....1B}. Any CDS would have to arise interior to this radius. Therefore, we believe that the IR emission most likely arises from pre-existing dust grains surrounding SN 2010jl heated by the initial flash. 

As mentioned above, pre-explosion IRAC images do exist for UGC 5189A, but we are unable to detect the CSM shell surrounding the progenitor.  This is not surprising if we assume either a red supergiant (RSG, L = 5 $\times$ 10$^{4}$ L$_{\sun}$, T = 3500K) or an LBV (L = 1 $\times$ 10$^{6}$ L$_{\sun}$, T = 10000 K) as a progenitor star, 48.9 Mpc away with dust at a distance of  6 $\times$ 10$^{17}$ cm,  the total flux contribution from both the star and the dust in the pre-explosion image in 3.6 and 4.6 $\mu$m is less than 0.5 $\mu$Jy.  

In order to estimate the pre-existing dust mass in SN 2010jl, we used our radiative transfer code MOCASSIN, a three-dimensional Monte Carlo radiative-transfer code that is capable of modeling non-spherical geometries, clumpy density distributions, and non-central energy sources \citep{2003MNRAS.340.1153E,2005MNRAS.362.1038E,2008ApJS..175..534E}. As an initial fit, we used a ``smooth" model, in which the dust is uniformly distributed throughout a spherical shell according to a r$^{-2}$ density profile, and constrained within a shell with size R$_{in}$ and R$_{out}$. We used a standard MRN grain size distribution of $a^{-3.5}$ between 0.005 and 0.05 $\mu$m \citep{1977ApJ...217..425M}, as was used in previous modeling done by \cite{2007ApJ...665..608M}, \cite{2009ApJ...704..306K}, and \cite{2011ApJ...731...47A}. Due to the early epoch of the observations, we assume the SN is a point source at the center of a spherical shell.

We used version 2.02.67 of MOCASSIN, which includes the
 capacity to account for light travel time effects (see \citet{2010MNRAS.403..474W}  for details), to calculate the emission from only that portion of the dust illuminated after 90 days (Figure \ref{fig:echo}). When light travel times are taken into account, after 90 days only about 5$\%$ of a spherical shell with  R$_{in}$ = 6  $\times$ 10$^{17}$ cm and R$_{out}$ = 1.4  $\times$ 10$^{18}$ cm would be illuminated as seen from Earth (see Figure 4). Applying this constraint, spherical dust geometries could provide a
 good match to the observed day 90 SED, with a total dust mass of $\sim$0.6 M$_{\sun}$, but gave a line-of-sight optical depth of $\sim$6.  This was problematic since previous papers from \citet{2011A&A...527L...6P} and \citep{2011ApJ...732...63S} estimated a Galactic foreground reddening of  E(B-V) = 0.027 for SN 2010jl and no reddening from the host galaxy. SN 2010jl is a good match to SN 2006tf, which also had little host galaxy reddening  \citep{2011ApJ...732...63S}.   \citet{2011A&A...527L...6P}, using both the continuum of the optical spectra and equivalent widths of the Na I D lines, also found no host galaxy reddening.  Therefore, a geometry other than a spherical shell, such as an inclined torus or bi-polar lobes, needs to be considered in order to allow low optical depth along the line of sight, but with sufficient dust mass to account for a CSM envelope with the high IR excess of SN 2010jl.

For the torus models we used a uniform dust density with a distance of 1.2 light years (6 $\times$ 10$^{17}$ cm) and a tube radius of 0.5 light years (4.7 $\times$ 10$^{17}$ cm) \citep{2007MNRAS.375..753E}. A torus with zero inclination (in plane of the sky) would not have been illuminated by the supernova by day 90. An inclination of at least 45$^{\circ}$ is required for IR emission to be present on day 90.  Figure \ref{fig:SED}, shows the fits for the 60$^{\circ}$ inclination.  We used both 100$\%$ amorphous carbon (AC) compositions and 100$\%$ silicate (Si) compositions, since the lack of data beyond 4.5 $\mu$m, makes it impossible to differentiate between silicate and AC grains. We find an inner radius of 6  $\times$ 10$^{17}$ cm also provided the best fit to the data, which is  actually larger (particularly for a carbon rich CSM) than the predicted evaporation radius discussed above. This suggests that dust was absent interior to r$_{evap}$ before the explosion, and that the pre-existing dust was actually formed by a progenitor eruption  as early as 300 years prior (if we assume an LBV eruption of $\sim$ 600 km s$^{-1}$, as was seen in the bipolar lobes of $\eta$ Car \citep{2003ApJ...586..432S}) or as late as 20,000 years if we assume an RSG progenitor with a constant wind of 10 km s$^{-1}$.

The 45$^{\circ}$ inclined torus, although having a zero optical depth underestimates the IR flux.  The 80$^{\circ}$ torus does a fairly good job fitting the data, but has an optical depth of 1.53 and 2.4 for the AC and Si models respectively, much too high for the reddening suggested by the early time optical spectra. The 60$^{\circ}$ torus includes both a zero optical depth and an acceptable fit to the data, making it the most likely inclination of the surrounding CSM torus.  With only the two mid-IR points the exact inclination is not very well constrained, but an inclination of 60 - 80$^{\circ}$ to the plane of the sky surrounding SN 2010jl is consistent with the data.  This model (see input parameters listed in Tables 4 and 5), implies that between 0.03 and 0.35 M$_{\sun}$ of pre-existing dust surrounds SN 2010jl. This would indicate total masses of CSM between 3- 35 M$_{\sun}$, assuming a gas to dust ratio of 100 and the possible inclinations and dust compositions.  Although this is a large amount, measurements of dust around Eta Carinae, an extensively studied LBV in our galaxy, yield dust values between 0.4 - 0.7 M$_{\sun}$, \citep{2006MNRAS.372.1133G,2010MNRAS.401L..48G} making it plausible that previous mass loss events could have ejected this much dust around SN 2010jl.  Like the geometry of $\eta$ Carinae, we cannot rule out the dust being contained in bipolar lobes and not in a torus. However, observations at later epochs will provide stronger constraints on the
 geometry of the dusty CSM which we have shown to exist around SN 2010jl.

\section{Summary}
This paper presents the results of SED modeling of SN 2010jl $\sim$ 90 days post-explosion.  Using optical, near- and mid-IR observations of SN 2010jl from day 90 we have estimated a pre-existing dust mass between M$_{d}$ = 0.03 - 0.35 M$_{\sun}$, with the higher dust masses corresponding to an oxygen-rich CSM. This dust is likely located in a torus inclined between 60-80$^{\circ}$ some  6 $\times$ 10$^{17}$ cm away from the SN. The inclined torus is required by the conclusion of  \citet{2011A&A...527L...6P} and \citep{2011ApJ...732...63S} that there is very little local reddening around SN 2010jl, while at the same time allowing the existence of substantial amounts of dust needed to create the observed IR excess. The lack of dust between the estimated evaporation radius and the inner dust radius suggests that this torus of pre-existing dust was created in a mass loss episode likely 300 - 2000 years prior, assuming an LBV progenitor with expansion velocities between 100-600 km s$^{-1}$. As many LBVs can have bipolar or toroidal nebulae and massive dust shells, it seems very likely that the progenitor was an LBV.  This agrees with the conclusions of  \cite{2011ApJ...732...63S}, who suggests the progenitor was a massive LBV star.  If the inner edge of the CSM dust is 6 $\times$ 10$^{17}$ cm away we estimate that the contribution to the elevated IR flux from pre-existing dust will last at least for 1.2 years, and that it in fact will increase as more and more of the torus is illuminated.  Continued monitoring of this object is crucial in order to detect the formation of new dust, and to measure how much dust is created.

\acknowledgments
We would like to thank the anonymous referee for the valuable suggestions that have improved this paper. This work has been supported by NSF grant AST-0707691 and NASA GSRP grant  NNX08AV36H. This work was supported by Spitzer Space Telescope RSA 1415602 and RSA 1346842,  issued by JPL/Caltech. The standard data acquisition has been supported by NSF grants AST-0503871 and AST-0803158 to A. U. Landolt. We acknowledge with thanks the variable star observations from the AAVSO International Database contributed by observers worldwide and used in this research. DLW acknowledges support from a Natural Sciences and Engineering Research Council of Canada (NSERC) Discovery Grant


\begin{thebibliography}{41}
\expandafter\ifx\csname natexlab\endcsname\relax\def\natexlab#1{#1}\fi

\bibitem[{{Andrews} {et~al.}(2010){Andrews}, {Gallagher}, {Clayton},
  {Sugerman}, {Chatelain}, {Clem}, {Welch}, {Barlow}, {Ercolano}, {Fabbri},
  {Wesson}, \& {Meixner}}]{2010ApJ...715..541A}
{Andrews}, J.~E., {Gallagher}, J.~S., {Clayton}, G.~C., {Sugerman}, B.~E.~K.,
  {Chatelain}, J.~P., {Clem}, J., {Welch}, D.~L., {Barlow}, M.~J., {Ercolano},
  B., {Fabbri}, J., {Wesson}, R., \& {Meixner}, M. 2010, \apj, 715, 541

\bibitem[{{Andrews} {et~al.}(2011){Andrews}, {Sugerman}, {Clayton},
  {Gallagher}, {Barlow}, {Clem}, {Ercolano}, {Fabbri}, {Meixner}, {Otsuka},
  {Welch}, \& {Wesson}}]{2011ApJ...731...47A}
{Andrews}, J.~E., {Sugerman}, B.~E.~K., {Clayton}, G.~C., {Gallagher}, J.~S.,
  {Barlow}, M.~J., {Clem}, J., {Ercolano}, B., {Fabbri}, J., {Meixner}, M.,
  {Otsuka}, M., {Welch}, D.~L., \& {Wesson}, R. 2011, \apj, 731, 47

\bibitem[{{Benetti} {et~al.}(2010){Benetti}, {Bufano}, {Vinko}, {Marion},
  {Pritchard}, {Wheeler}, {Chatzopoulos}, \& {Shetrone}}]{2010CBET.2536....1B}
{Benetti}, S., {Bufano}, F., {Vinko}, J., {Marion}, G.~H., {Pritchard}, T.,
  {Wheeler}, J.~C., {Chatzopoulos}, E., \& {Shetrone}, M. 2010, Central Bureau
  Electronic Telegrams, 2536, 1

\bibitem[{{Cherchneff} \& {Dwek}(2010)}]{2010ApJ...713....1C}
{Cherchneff}, I. \& {Dwek}, E. 2010, \apj, 713, 1

\bibitem[{{Chevalier} \& {Fransson}(1994)}]{1994ApJ...420..268C}
{Chevalier}, R.~A. \& {Fransson}, C. 1994, \apj, 420, 268

\bibitem[{{Dwek}(1983)}]{1983ApJ...274..175D}
{Dwek}, E. 1983, \apj, 274, 175

\bibitem[{{Dwek}(1985)}]{1985ApJ...297..719D}
---. 1985, \apj, 297, 719

\bibitem[{{Elmhamdi} {et~al.}(2003){Elmhamdi}, {Danziger}, {Chugai},
  {Pastorello}, {Turatto}, {Cappellaro}, {Altavilla}, {Benetti}, {Patat}, \&
  {Salvo}}]{2003MNRAS.338..939E}
{Elmhamdi}, A., {Danziger}, I.~J., {Chugai}, N., {Pastorello}, A., {Turatto},
  M., {Cappellaro}, E., {Altavilla}, G., {Benetti}, S., {Patat}, F., \&
  {Salvo}, M. 2003, \mnras, 338, 939

\bibitem[{{Ercolano} {et~al.}(2005){Ercolano}, {Barlow}, \&
  {Storey}}]{2005MNRAS.362.1038E}
{Ercolano}, B., {Barlow}, M.~J., \& {Storey}, P.~J. 2005, \mnras, 362, 1038

\bibitem[{{Ercolano} {et~al.}(2007){Ercolano}, {Barlow}, \&
  {Sugerman}}]{2007MNRAS.375..753E}
{Ercolano}, B., {Barlow}, M.~J., \& {Sugerman}, B.~E.~K. 2007, \mnras, 375, 753

\bibitem[{{Ercolano} {et~al.}(2003){Ercolano}, {Morisset}, {Barlow}, {Storey},
  \& {Liu}}]{2003MNRAS.340.1153E}
{Ercolano}, B., {Morisset}, C., {Barlow}, M.~J., {Storey}, P.~J., \& {Liu}, X.
  2003, \mnras, 340, 1153

\bibitem[{{Ercolano} {et~al.}(2008){Ercolano}, {Young}, {Drake}, \&
  {Raymond}}]{2008ApJS..175..534E}
{Ercolano}, B., {Young}, P.~R., {Drake}, J.~J., \& {Raymond}, J.~C. 2008,
  \apjs, 175, 534

\bibitem[{{Fox} {et~al.}(2009){Fox}, {Skrutskie}, {Chevalier}, {Kanneganti},
  {Park}, {Wilson}, {Nelson}, {Amirhadji}, {Crump}, {Hoeft}, {Provence},
  {Sargeant}, {Sop}, {Tea}, {Thomas}, \& {Woolard}}]{2009ApJ...691..650F}
{Fox}, O., {Skrutskie}, M.~F., {Chevalier}, R.~A., {Kanneganti}, S., {Park},
  C., {Wilson}, J., {Nelson}, M., {Amirhadji}, J., {Crump}, D., {Hoeft}, A.,
  {Provence}, S., {Sargeant}, B., {Sop}, J., {Tea}, M., {Thomas}, S., \&
  {Woolard}, K. 2009, \apj, 691, 650

\bibitem[{{Fox} {et~al.}(2011){Fox}, {Chevalier}, {Skrutskie}, {Soderberg},
  {Filippenko}, {Ganeshalingam}, {Silverman}, {Smith}, \&
  {Steele}}]{2011arXiv1104.5012F}
{Fox}, O.~D., {Chevalier}, R.~A., {Skrutskie}, M.~F., {Soderberg}, A.~M.,
  {Filippenko}, A.~V., {Ganeshalingam}, M., {Silverman}, J.~M., {Smith}, N., \&
  {Steele}, T.~N. 2011, ArXiv e-prints

\bibitem[{{Gal-Yam} \& {Leonard}(2009)}]{2009Natur.458..865G}
{Gal-Yam}, A. \& {Leonard}, D.~C. 2009, \nat, 458, 865

\bibitem[{{Gal-Yam} {et~al.}(2007){Gal-Yam}, {Leonard}, {Fox}, {Cenko},
  {Soderberg}, {Moon}, {Sand}, {Li}, {Filippenko}, {Aldering}, \&
  {Copin}}]{2007ApJ...656..372G}
{Gal-Yam}, A., {Leonard}, D.~C., {Fox}, D.~B., {Cenko}, S.~B., {Soderberg},
  A.~M., {Moon}, D.-S., {Sand}, D.~J., {Li}, W., {Filippenko}, A.~V.,
  {Aldering}, G., \& {Copin}, Y. 2007, \apj, 656, 372

\bibitem[{{Gomez} {et~al.}(2010){Gomez}, {Vlahakis}, {Stretch}, {Dunne},
  {Eales}, {Beelen}, {Gomez}, \& {Edmunds}}]{2010MNRAS.401L..48G}
{Gomez}, H.~L., {Vlahakis}, C., {Stretch}, C.~M., {Dunne}, L., {Eales}, S.~A.,
  {Beelen}, A., {Gomez}, E.~L., \& {Edmunds}, M.~G. 2010, \mnras, 401, L48

\bibitem[{{Gomez (N{\'e}e Morgan)} {et~al.}(2006){Gomez (N{\'e}e Morgan)},
  {Dunne}, {Eales}, \& {Edmunds}}]{2006MNRAS.372.1133G}
{Gomez (N{\'e}e Morgan)}, H.~L., {Dunne}, L., {Eales}, S.~A., \& {Edmunds},
  M.~G. 2006, \mnras, 372, 1133

\bibitem[{{Kiewe} {et~al.}(2010){Kiewe}, {Gal-Yam}, {Arcavi}, {Leonard},
  {Emilio Enriquez}, {Cenko}, {Fox}, {Moon}, {Sand}, \&
  {Soderberg}}]{2010arXiv1010.2689K}
{Kiewe}, M., {Gal-Yam}, A., {Arcavi}, I., {Leonard}, D.~C., {Emilio Enriquez},
  J., {Cenko}, S.~B., {Fox}, D.~B., {Moon}, D., {Sand}, D.~J., \& {Soderberg},
  A.~M. 2010, ArXiv e-prints

\bibitem[{{Kotak} {et~al.}(2009){Kotak}, {Meikle}, {Farrah}, {Gerardy},
  {Foley}, {Van Dyk}, {Fransson}, {Lundqvist}, {Sollerman}, {Fesen},
  {Filippenko}, {Mattila}, {Silverman}, {Andersen}, {H{\"o}flich}, {Pozzo}, \&
  {Wheeler}}]{2009ApJ...704..306K}
{Kotak}, R., {Meikle}, W.~P.~S., {Farrah}, D., {Gerardy}, C.~L., {Foley},
  R.~J., {Van Dyk}, S.~D., {Fransson}, C., {Lundqvist}, P., {Sollerman}, J.,
  {Fesen}, R., {Filippenko}, A.~V., {Mattila}, S., {Silverman}, J.~M.,
  {Andersen}, A.~C., {H{\"o}flich}, P.~A., {Pozzo}, M., \& {Wheeler}, J.~C.
  2009, \apj, 704, 306

\bibitem[{{Landolt}(2009)}]{2009AJ....137.4186L}
{Landolt}, A.~U. 2009, \aj, 137, 4186

\bibitem[{{Li} {et~al.}(2011){Li}, {Leaman}, {Chornock}, {Filippenko},
  {Poznanski}, {Ganeshalingam}, {Wang}, {Modjaz}, {Jha}, {Foley}, \&
  {Smith}}]{2011MNRAS.412.1441L}
{Li}, W., {Leaman}, J., {Chornock}, R., {Filippenko}, A.~V., {Poznanski}, D.,
  {Ganeshalingam}, M., {Wang}, X., {Modjaz}, M., {Jha}, S., {Foley}, R.~J., \&
  {Smith}, N. 2011, \mnras, 412, 1441

\bibitem[{{Mathis} {et~al.}(1977){Mathis}, {Rumpl}, \&
  {Nordsieck}}]{1977ApJ...217..425M}
{Mathis}, J.~S., {Rumpl}, W., \& {Nordsieck}, K.~H. 1977, \apj, 217, 425

\bibitem[{{Meikle} {et~al.}(2007){Meikle}, {Mattila}, {Pastorello}, {Gerardy},
  {Kotak}, {Sollerman}, {Van Dyk}, {Farrah}, {Filippenko}, {H{\"o}flich},
  {Lundqvist}, {Pozzo}, \& {Wheeler}}]{2007ApJ...665..608M}
{Meikle}, W.~P.~S., {Mattila}, S., {Pastorello}, A., {Gerardy}, C.~L., {Kotak},
  R., {Sollerman}, J., {Van Dyk}, S.~D., {Farrah}, D., {Filippenko}, A.~V.,
  {H{\"o}flich}, P., {Lundqvist}, P., {Pozzo}, M., \& {Wheeler}, J.~C. 2007,
  \apj, 665, 608

\bibitem[{{Meixner} {et~al.}(2008){Meixner}, {Smee}, {Doering}, {Barkhouser},
  {Miller}, {Orndoff}, {Knezek}, {Churchwell}, {Scharfstein}, {Percival},
  {Mills}, \& {Corson}}]{2008SPIE.7014E..96M}
{Meixner}, M., {Smee}, S., {Doering}, R.~L., {Barkhouser}, R.~H., {Miller}, T.,
  {Orndoff}, J., {Knezek}, P., {Churchwell}, E., {Scharfstein}, G., {Percival},
  J., {Mills}, D., \& {Corson}, C. 2008, in Presented at the Society of
  Photo-Optical Instrumentation Engineers (SPIE) Conference, Vol. 7014, Society
  of Photo-Optical Instrumentation Engineers (SPIE) Conference Series

\bibitem[{{Newton} \& {Puckett}(2010)}]{2010CBET.2532....1N}
{Newton}, J. \& {Puckett}, T. 2010, Central Bureau Electronic Telegrams, 2532,
  1

\bibitem[{{Patat} {et~al.}(2011){Patat}, {Taubenberger}, {Benetti},
  {Pastorello}, \& {Harutyunyan}}]{2011A&A...527L...6P}
{Patat}, F., {Taubenberger}, S., {Benetti}, S., {Pastorello}, A., \&
  {Harutyunyan}, A. 2011, \aap, 527, L6+

\bibitem[{{Pozzo} {et~al.}(2004){Pozzo}, {Meikle}, {Fassia}, {Geballe},
  {Lundqvist}, {Chugai}, \& {Sollerman}}]{2004MNRAS.352..457P}
{Pozzo}, M., {Meikle}, W.~P.~S., {Fassia}, A., {Geballe}, T., {Lundqvist}, P.,
  {Chugai}, N.~N., \& {Sollerman}, J. 2004, \mnras, 352, 457

\bibitem[{{Schlegel} {et~al.}(1998){Schlegel}, {Finkbeiner}, \&
  {Davis}}]{1998ApJ...500..525S}
{Schlegel}, D.~J., {Finkbeiner}, D.~P., \& {Davis}, M. 1998, \apj, 500, 525

\bibitem[{{Schlegel}(1990)}]{1990MNRAS.244..269S}
{Schlegel}, E.~M. 1990, \mnras, 244, 269

\bibitem[{{Smartt} {et~al.}(2009){Smartt}, {Eldridge}, {Crockett}, \&
  {Maund}}]{2009MNRAS.395.1409S}
{Smartt}, S.~J., {Eldridge}, J.~J., {Crockett}, R.~M., \& {Maund}, J.~R. 2009,
  \mnras, 395, 1409

\bibitem[{{Smith} {et~al.}(2010){Smith}, {Chornock}, {Silverman}, {Filippenko},
  \& {Foley}}]{2010ApJ...709..856S}
{Smith}, N., {Chornock}, R., {Silverman}, J.~M., {Filippenko}, A.~V., \&
  {Foley}, R.~J. 2010, \apj, 709, 856

\bibitem[{{Smith} {et~al.}(2003){Smith}, {Davidson}, {Gull}, {Ishibashi}, \&
  {Hillier}}]{2003ApJ...586..432S}
{Smith}, N., {Davidson}, K., {Gull}, T.~R., {Ishibashi}, K., \& {Hillier},
  D.~J. 2003, \apj, 586, 432

\bibitem[{{Smith} {et~al.}(2008){Smith}, {Foley}, \&
  {Filippenko}}]{2008ApJ...680..568S}
{Smith}, N., {Foley}, R.~J., \& {Filippenko}, A.~V. 2008, \apj, 680, 568

\bibitem[{{Smith} {et~al.}(2011{\natexlab{a}}){Smith}, {Li}, {Filippenko}, \&
  {Chornock}}]{2011MNRAS.412.1522S}
{Smith}, N., {Li}, W., {Filippenko}, A.~V., \& {Chornock}, R.
  2011{\natexlab{a}}, \mnras, 412, 1522

\bibitem[{{Smith} {et~al.}(2011{\natexlab{b}}){Smith}, {Li}, {Miller},
  {Silverman}, {Filippenko}, {Cuillandre}, {Cooper}, {Matheson}, \& {Van
  Dyk}}]{2011ApJ...732...63S}
{Smith}, N., {Li}, W., {Miller}, A.~A., {Silverman}, J.~M., {Filippenko},
  A.~V., {Cuillandre}, J.-C., {Cooper}, M.~C., {Matheson}, T., \& {Van Dyk},
  S.~D. 2011{\natexlab{b}}, \apj, 732, 63

\bibitem[{{Smith} {et~al.}(2009){Smith}, {Silverman}, {Chornock}, {Filippenko},
  {Wang}, {Li}, {Ganeshalingam}, {Foley}, {Rex}, \&
  {Steele}}]{2009ApJ...695.1334S}
{Smith}, N., {Silverman}, J.~M., {Chornock}, R., {Filippenko}, A.~V., {Wang},
  X., {Li}, W., {Ganeshalingam}, M., {Foley}, R.~J., {Rex}, J., \& {Steele},
  T.~N. 2009, \apj, 695, 1334

\bibitem[{{Stoll} {et~al.}(2010){Stoll}, {Prieto}, {Stanek}, {Pogge},
  {Szczygiel}, {Pojmanski}, {Antognini}, \& {Yan}}]{2010arXiv1012.3461S}
{Stoll}, R., {Prieto}, J.~L., {Stanek}, K.~Z., {Pogge}, R.~W., {Szczygiel},
  D.~M., {Pojmanski}, G., {Antognini}, J., \& {Yan}, H. 2010, ArXiv e-prints

\bibitem[{{Sugerman} {et~al.}(2006){Sugerman}, {Ercolano}, {Barlow}, {Tielens},
  {Clayton}, {Zijlstra}, {Meixner}, {Speck}, {Gledhill}, {Panagia}, {Cohen},
  {Gordon}, {Meyer}, {Fabbri}, {Bowey}, {Welch}, {Regan}, \&
  {Kennicutt}}]{2006Sci...313..196S}
{Sugerman}, B.~E.~K., {Ercolano}, B., {Barlow}, M.~J., {Tielens}, A.~G.~G.~M.,
  {Clayton}, G.~C., {Zijlstra}, A.~A., {Meixner}, M., {Speck}, A., {Gledhill},
  T.~M., {Panagia}, N., {Cohen}, M., {Gordon}, K.~D., {Meyer}, M., {Fabbri},
  J., {Bowey}, J.~E., {Welch}, D.~L., {Regan}, M.~W., \& {Kennicutt}, R.~C.
  2006, Science, 313, 196

\bibitem[{{Wesson} {et~al.}(2010){Wesson}, {Barlow}, {Ercolano}, {Andrews},
  {Clayton}, {Fabbri}, {Gallagher}, {Meixner}, {Sugerman}, {Welch}, \&
  {Stock}}]{2010MNRAS.403..474W}
{Wesson}, R., {Barlow}, M.~J., {Ercolano}, B., {Andrews}, J.~E., {Clayton},
  G.~C., {Fabbri}, J., {Gallagher}, J.~S., {Meixner}, M., {Sugerman}, B.~E.~K.,
  {Welch}, D.~L., \& {Stock}, D.~J. 2010, \mnras, 403, 474

\bibitem[{{Yamanaka} {et~al.}(2010){Yamanaka}, {Okushima}, {Arai}, {Sasada}, \&
  {Sato}}]{2010CBET.2539....1Y}
{Yamanaka}, M., {Okushima}, T., {Arai}, A., {Sasada}, M., \& {Sato}, H. 2010,
  Central Bureau Electronic Telegrams, 2539, 1

\end{thebibliography}

\clearpage

\begin{table*}[h!]
\caption{Tertiary BVRI Standards for UGC 5189}
\centering
\begin{tabular}{cccccc}
\hline
\hline
Star&U&B&V&R&I\\
\hline
A&20.553 $\pm$ 0.1351 &19.611 $\pm$ 0.0453&18.486 $\pm$ 0.0180&17.765 $\pm$ 0.0148&17.152 $\pm$ 0.0366\\
B&19.087 $\pm$ 0.0429&18.468 $\pm$ 0.0176&17.513 $\pm$ 0.0123&16.963 $\pm$ 0.0084&16.488 $\pm$ 0.0200\\
C&14.929 $\pm$ 0.0138&14.677 $\pm$ 0.0036&13.944 $\pm$ 0.0049&13.521 $\pm$ 0.0048&13.131 $\pm$ 0.0038\\
D&19.807 $\pm$ 0.3422&19.470 $\pm$ 0.0508&18.164 $\pm$ 0.0138&17.313 $\pm$ 0.0167&16.606 $\pm$ 0.0153\\
E&19.967 $\pm$ 0.0636&18.751 $\pm$ 0.0537&17.189 $\pm$ 0.0065&16.062 $\pm$ 0.0111&14.685 $\pm$ 0.0052\\
F&19.888 $\pm$ 0.0995&19.456 $\pm$ 0.0448&18.212 $\pm$ 0.0201&17.511 $\pm$ 0.0154&16.823 $\pm$ 0.0450\\
G&16.986 $\pm$ 0.0157&16.323 $\pm$ 0.0060&15.405 $\pm$ 0.0027&14.877 $\pm$ 0.0059&14.427 $\pm$ 0.0055\\
H&18.965 $\pm$ 0.0834&19.022 $\pm$ 0.0597&18.310 $\pm$ 0.0162&17.848$\pm$ 0.0220&17.504 $\pm$ 0.0274\\
I& 20.439 $\pm$ 0.3093&19.566 $\pm$ 0.1121&18.433 $\pm$ 0.0174&17.685 $\pm$ 0.0211& 17.053 $\pm$ 0.0334\\
\hline
\end{tabular}
\centering
\end{table*}
\label{tab:standards}
\clearpage

\begin{table*}[h]
\caption{BVRI observations of SN 2010jl}
\centering
\begin{tabular}{ccccccc}
\hline
\hline
JD&Age&B&V&R&I&Telescope/Instrument\\
\hline
2455535&55&14.32 &13.94 &13.51 &13.32& KPNO/1.2m\\
&&$\pm$ 0.01&$\pm$ 0.01& $\pm$ 0.01& $\pm$ 0.01&\\
2455629&149&15.16 &14.69 &14.07 &14.08& KPNO/1.2m\\
&&$\pm$ 0.01&$\pm$ 0.02& $\pm$ 0.01& $\pm$ 0.01&\\
2455656&176&15.20 &14.76 &14.09 &14.16& KPNO/1.2m\\
&&$\pm$ 0.03&$\pm$ 0.01& $\pm$ 0.01& $\pm$ 0.01&\\
\hline
\end{tabular}
\centering
\label{tab:Visphot}
\end{table*}

\begin{table*}[h]
\caption{IR observations of SN 2010jl}
\centering
\begin{tabular}{ccccccccc}
\hline
\hline
JD&Age&J&H&K$_{s}$&3.6 $\mu$m&4.5 $\mu$m&Telescope/Instrument\\
\hline
2455565&90&&&&4.04 mJy&4.52 mJy& Spitzer/IRAC\\
&&&&&$\pm$ 0.14& $\pm$ 0.18&\\
2455583&108&13.41&13.09&13.75&&&WIYN/WHIRC\\
&&$\pm$ 0.2&$\pm$ 0.2&$\pm$ 0.2&&&\\
\hline
\end{tabular}
\centering
\label{tab:IRphot}
\end{table*}

\begin{table*}[h]
\caption {Monte Carlo Radiative Transfer Torus Models for 100$\%$ Amorphous Carbon}
\centering
\begin{tabular}{cccccccccc}
\hline
\hline
Inclination  & T$_{ej}$ (K) & R$_{in}$ (cm) & R$_{out}$(cm) & L$_{tot.}$ (L$_{\odot}$) & $\tau$$_{v}$ &  M$_{d}$ (M$_{\odot}$)   \\
\hline
45$^{\circ}$ & 7500 &6e17 &1.4e18  & 5.5e9 & 0.0 &0.12  \\
60$^{\circ}$ &  7500 &6e17 &1.4e18  & 5.5e9 & 0.0 & 0.05 \\
80$^{\circ}$ &  7500 &6e17 &1.4e18  & 5.5e9 & 1.5 & 0.03\\
\hline
\end{tabular}
\centering
\label{tab:Carbon}
\end{table*}

\begin{table*}[h]
\caption {Monte Carlo Radiative Transfer Torus Models for 100$\%$ Silicates}
\centering
\begin{tabular}{cccccccccc}
\hline
\hline
Inclination  & T$_{ej}$ (K) & R$_{in}$ (cm) & R$_{out}$(cm) & L$_{tot.}$ (L$_{\odot}$) & $\tau$$_{v}$ &  M$_{d}$ (M$_{\odot}$)   \\
\hline
45$^{\circ}$ & 7500 &6e17 &1.4e18  & 2.6e10 & 0.0 &0.7  \\
60$^{\circ}$ &  7500 &6e17 &1.4e18  & 2.6e10 & 0.0 & 0.35 \\
80$^{\circ}$ &  7500 &6e17 &1.4e18  & 2.6e10 & 2.4 & 0.27\\
\hline
\end{tabular}
\centering
\label{tab:Silicate}
\end{table*}


\begin{figure}[h]
   \centering
   \includegraphics[width=4in]{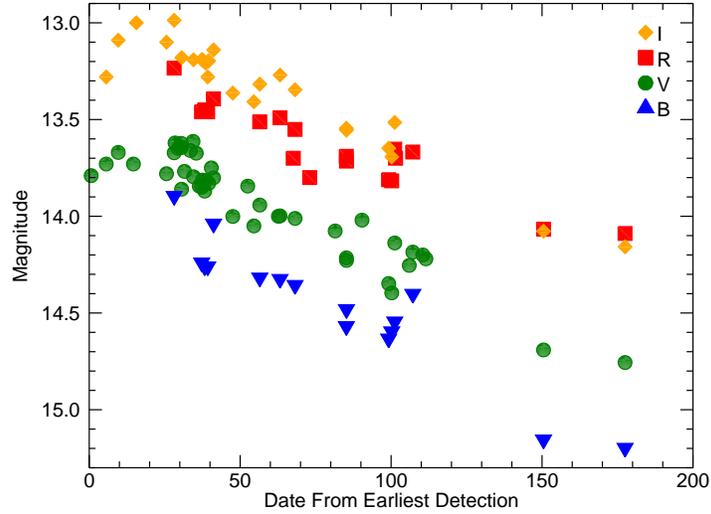} 
   \caption{ Optical lightcurve of SN 2010jl.  Data are from AAVSO, Stoll et. al 2010, and new data presented in Table \ref{tab:Visphot} in this paper.}
   \label{fig:LC}
\end{figure}

\begin{figure}[htbp]
   \centering
   \includegraphics[angle=90,width=4.5in]{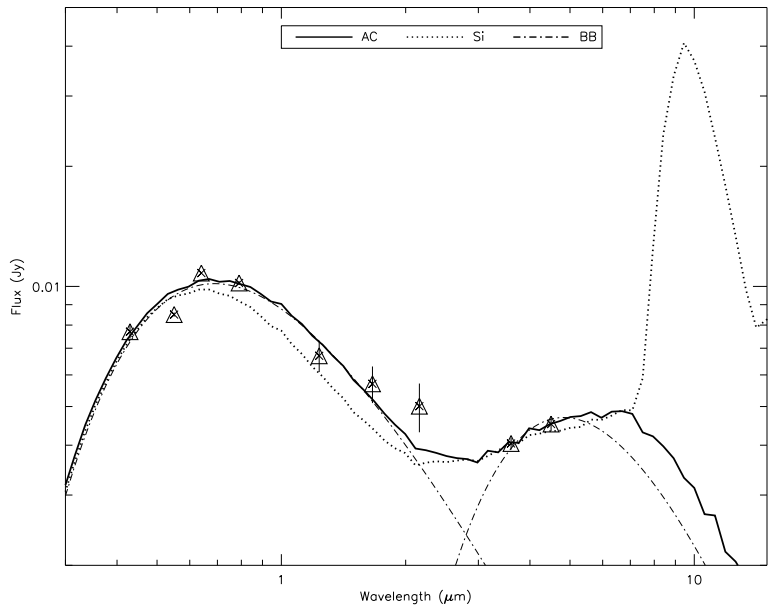}
   \caption{SED of SN 2010jl on $\sim$ day 90.  IR photometry is listed in Table \ref{tab:IRphot}, and optical photometry was obtained from the AAVSO, specifically contributor Etienne Morelle.  Dashed dotted lines are the blackbody fits to the data. The solid line indicates the 100$\%$ amorphous carbon inclined at 60$^{\circ}$ MOCASSIN torus model fit, the dotted line is the same except for the composition being 100$\%$ silicates.  As the figure shows, without observations longward of 4.6 $\mu$m, we cannot accurately constrain the dust composition.}
   \label{fig:SED}
\end{figure}

\begin{figure}[h] 
   \centering
   \includegraphics[width=4in]{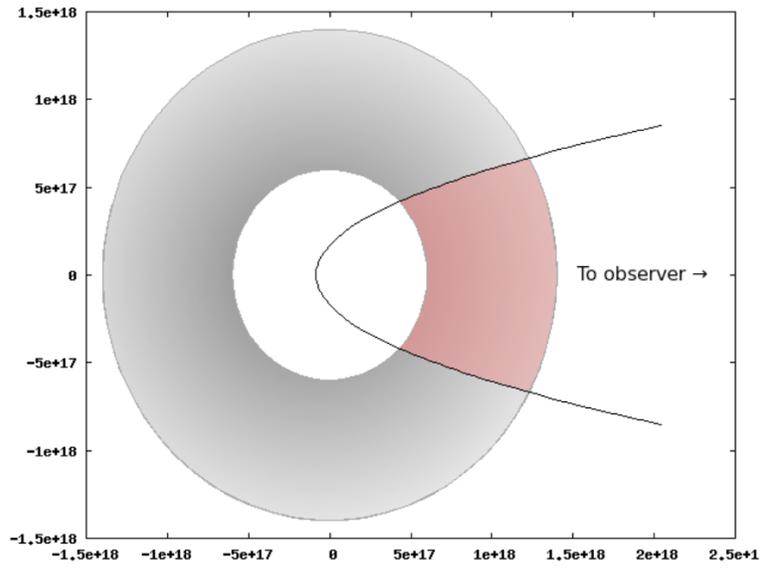} 
   \caption{Schematic of the fraction of the CSM ejecta which will have been lit when light travel times are taken into consideration.  For the example above, at 90 days, only about 5$\%$ would have been illuminated.   }
   \label{fig:echo}
\end{figure}

\end{document}